\documentclass[12pt]{article}
\usepackage{amssymb}
\usepackage{amsmath}
\usepackage{hyperref}
\usepackage{graphicx}
\usepackage{placeins}
\usepackage{bm}
\usepackage{latexsym}
\begin{document}
\title{\bf{AdS Black Hole with Cylindrical Symmetry}}
\author{Mehdi Sadeghi\thanks{Corresponding author: Email: mehdi.sadeghi@abru.ac.ir} ,\hspace{2mm} Ramin Anvari Asl   \thanks{r.anvari11@gmail.com} ,\hspace{2mm} Mohammad Shamseh \thanks{mohamadshamseh0@gmail.com}\\
		 {\small {\em  Department of Physics, School of Sciences,}}\\
		 {\small {\em  Ayatollah Boroujerdi University, Boroujerd, Iran}}\\
   }
\date{\today}
\maketitle

\abstract{ In this paper, we consider Einstein-Hilbert gravity in the presence of cosmological constant with cylindrical symmetry to introduce the black hole solution of this model. Here, we solve the Einstein's vacuum field equation, and then  we calculate the appropriate metric for this problem.}\\


 \noindent \textbf{Keywords:} Black hole; Einstein's field equations; Cylindrical symmetry

 \section{Introduction} \label{intro}

  General relativity is a theory of gravity formulated by Albert Einstein in 1915. In this theory, gravity is massless and spin-2 \cite{Einstein:1914bt}-\cite{Einstein:1915by}. General relativity is a gauge theory and it is invariant under diffeomorphism. The presence of matter induces gravity by curving the space-time in general relativity. Solving Einstein's field equations in the different situations is an interesting subject. For example, massive gravity\cite{deRham:2010kj}, higher derivative gravity\cite{Sadeghi:2015vaa}-\cite{Parvizi:2017boc} and Rastall gravity\cite{Sadeghi:2018vrf}, as modifications of gravity, were introduced for explanation of dark matter and dark energy.\\  
Levi-Civita \cite{Levi-Civita} and Weyl \cite{Weyl:1917gp} studied the cylindrical symmetry in general relativity for the first time and its black hole solution in an anti-de Sitter (AdS)
background has been studied in \cite{Lemos:1994xp}-\cite{Lemos:1997bd}. Cylindrical
symmetry is the most simple capable symmetry to produce gravitational waves.\\
In this paper, we have tried to solve Einstein's field equations for a mass line in AdS space-time with cylindrical symmetry.\\
The resulted state in a gravitational collapse of astrophysical massive objects is called the black hole. In other words,  a black hole is a solution of Einstein's equations with an event horizon. Event horizon is a null hyper-surface and one-dimensional. There is not a causal relation between the two sides of the event horizon.\\
Cosmic strings \cite{Alexander Vilenkin}, the role of conical singularities, space-time topological defects, gravitational waves and the observable acceleration of high
energy jets in galaxies \cite{Bronnikov:2019clf} are the concepts that could be explained  better  by the symmetric aspects of cylindrical space-time. Therefore, it motivates us to study the model with axial symmetry.
    \section{Einstein AdS Black Hole Solution with Cylindrical Symmetry  }
     \label{sec2}
\begin{figure}[h!]
	\centerline{\includegraphics[width=10cm]{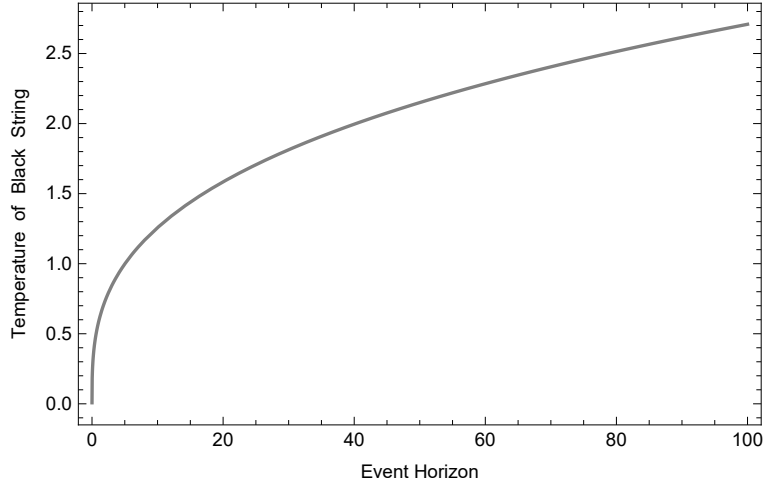}}
	\caption{$T-r_{+}$ diagram for $c=\frac{4}{3}$. \label{fig:1}}
\end{figure}

     The action for this model is written as follows,
\begin{equation}\label{action}
S=\frac{1}{2\kappa}\int{d^4x\sqrt{-g}(R-2\Lambda)},
\end{equation}
where, the first term, $R$, is the Ricci scalar and it represents the kinetic term of Einstein gravity. The second term ,$\Lambda$, is the cosmological constant expressed as,
\begin{equation}
\Lambda=-\frac{3}{l^2},
\end{equation}
 where $l$ is the radius of AdS space-time. Negative sign in $\Lambda$ guarantees that we are in AdS space-time.\\
We choose the following metric as an ansatz for the line element,
\begin{equation}\label{metric}
 ds^{2} =e^{\alpha(r)}dt^{2} -e^{\beta(r)}dr^{2} -r^{2}e^{\gamma(r)}d\varphi^{2}-e^{\gamma(r)}dz^{2}.
 \end{equation}
 The equations of motion can be derived by variation of the action (\ref{action}) with respect to the metric,
 \begin{equation}\label{EoM}
  {G_{\mu \nu }} + \Lambda {g_{\mu \nu }} = 0,
  \end{equation}
  where $G_{\mu \nu }= R_{\mu \nu}-\frac{1}{2}g_{\mu \nu}R$  is the Einstein tensor. We can write Eq.(\ref{EoM}) as the following,\\
  \begin{equation}\label{EoM1}
{R_{\mu \nu }} = \Lambda {g_{\mu \nu }}.
\end{equation}
The (t,t) component of Eq.(\ref{EoM1}) is written as the following by using the metric ansatz (\ref{metric}),
 \begin{equation}\label{tt}
\frac{1}{4 r}\Bigg(2r \alpha'(r) \gamma'(r)-r \alpha'(r) \beta'(r)+r \alpha'(r)^2+2 r \alpha''(r)+2 \alpha'(r)\Bigg)=-\Lambda\, e^{\beta(r)}.
\end{equation}
The (r,r) component of Eq.(\ref{EoM1}) is given by,
\begin{align}\label{rr}
&\frac{1}{4 r}\Bigg(-r \alpha'(r) \beta'(r)+r \alpha'(r)^2+2 r \alpha''(r)\nonumber\\&-2 r \beta'(r) \gamma'(r)-2 \beta'(r)+2 r \gamma'(r)^2+4 \gamma'(r)+4 r \gamma''(r)\Bigg)=-\Lambda\,  e^{\beta(r)}.
\end{align}
The $(\varphi,\varphi)$ component of Eq.(\ref{EoM1}) is as,
\begin{align}\label{phi-phi}
&\frac{1}{4 r}\Bigg(r \alpha'(r) \gamma'(r)+2 \alpha'(r)-r \beta'(r) \gamma'(r)\nonumber\\&-2 \beta'(r)+2 r \gamma'(r)^2+6 \gamma'(r)+2 r \gamma''(r)\Bigg)=-\Lambda\, e^{\beta(r)},
\end{align}
and the $(z,z)$ component of Eq.(\ref{EoM1}) becomes,
\begin{equation}\label{zz}
\frac{1}{4 r}\Bigg(r \alpha'(r) \gamma'(r)-r \beta'(r) \gamma'(r)+2 r \gamma'(r)^2+2 \gamma'(r)+2 r \gamma''(r)\Bigg)=-\Lambda\, e^{\beta(r)}.
\end{equation}
By subtracting Eq.(8) from Eq.(9), we will have,
\begin{equation}\label{8-9}
2\gamma' -  \beta' + \alpha' = 0.
\end{equation}
By plugging Eq.(\ref{8-9})  in Eq.(\ref{tt}), Eq.(\ref{phi-phi}) and Eq.(\ref{zz}) we obtain,
\begin{align}
 & (\alpha'' r + \alpha') = - 2\Lambda\, r\,  e^{\beta(r)}, \\
 & (\gamma'' r + \gamma') = - 2\Lambda\, r\,  e^{\beta(r)},\\
  &  \gamma'' r + 3 \gamma' - \beta' + \alpha' =  - 2 \Lambda\, r\,  e^{\beta(r)},
\end{align}
Eq.(13) is the same as Eq.(12) by using Eq.(10).\\
 By subtracting Eq.(11) from Eq.(12) we have,
\begin{equation}\label{dif}
(\alpha'' - \gamma') r + (\alpha' - \gamma') = 0.
\end{equation}
Eq.(\ref{dif}) can be written as,
\begin{equation}
\alpha' - \gamma' =  \frac{3 c}{r}, 
\end{equation}
by considering Eq.(\ref{8-9}) and Eq.(15) we get,
\begin{align}
&\gamma' = \frac{1}{3}\,\beta' - \frac{c}{r}, \\
&\alpha' = \frac{1}{3}\,\beta' + \frac{2 c}{r}.
\end{align}
By substituting Eq.(10) in Eq.(7) we will have,
\begin{equation}
\beta''\, r + \beta' = - 6 \Lambda\, r\,  e^{\beta}
\end{equation}
where, $\beta(r)$ is found as follows,
\begin{equation}
  \beta(r) =\ln\Bigg(- \frac{c_{1}^2}{12 \Lambda r^2\, {\cos}^2(\frac{1}{2}\,c_{1}(\ln r - c_{2}))}\Bigg) = \ln\Bigg(- \frac{c_{2}}{12 \Lambda r^2\, \cos^2 \theta}\Bigg).
\end{equation}
Here, $c_{1}$ and $c_{2}$ are constant and $\theta$ is given by,
\begin{equation}
  \theta = \frac{1}{2}\,c_{1}\,(\ln r - c_{2}),
\end{equation}
$\beta(r)$ should be satisfied in Eq.(\ref{rr}),
\begin{equation}
r \beta'' - \frac{1}{3} r\,{\beta'}^2 - \frac{1}{3}\,\beta' + \frac{3 c^2 - 2 c}{r} = -2 \Lambda\, r\, e^{\beta}.
\end{equation}
Eq.(21) can be simplified as,
\begin{align}
\Bigg(\frac{-c_{1}}{r}\,\tan\theta + \frac{2}{r} + \frac{c_{1}^2}{2 r}\,(1 + {\tan}^2\theta)\Bigg) - \frac{1}{3} r (\frac{c_{1}^2}{r^2}\,\tan^2\theta
+\frac{4}{r^2} - \frac{4 c_{1}}{r^2}\,\tan\theta) \nonumber\\
- \frac{1}{3}(\frac{c_{1}}{r}\,\tan\theta - \frac{2}{r})
+\frac{3 c^2 - 2 c}{r} = -2 \Lambda r(\frac{-c_{1}^2}{12 \Lambda r^2} (1 + \tan^2 \theta)).
\end{align}
We conclude the following relation,
\begin{equation}
c_{1}^2 + (3 c - 1)^2 + 3 = 0.
\end{equation}
By integrating of Eq.(16) and Eq.(17) with respects to $r$, we will have,
\begin{equation}
\gamma =\ln\Bigg(-\frac{\sqrt[3]{c_{1}^2} r^{-c}}{\sqrt[3]{12 \Lambda r^2 \cos^2\theta}}\Bigg),
\end{equation}
\begin{equation}
\alpha = \ln\Bigg(- \frac{r^{2 c} \sqrt[3]{c_{1}^2}}{\sqrt[3]{12 \Lambda r^2 \cos^2\theta}}\Bigg).
\end{equation}
Therefore, the metric of a mass line in the AdS space is as follows,
 \begin{equation}
  ds^{2} = \Bigg(\frac{-r^{2c_{2}} \sqrt[3]{c_{1}^2}}{\sqrt[3]{12 \Lambda r^2\, {\cos}^2 \theta}}\Bigg) dt^{2} - \Bigg(\frac{-c_{1}^2}{12 \Lambda r^2\, {\cos}^2 \theta}\Bigg) dr^{2} - \Bigg(\frac{-r^{-c_{2}} \sqrt[3]{c_{1}^2}}{\sqrt[3]{12 \Lambda r^2\, {\cos}^2} \theta}\Bigg)(r^2 d\varphi^{2} + dz^{2}),
 \end{equation}
where $ c_{1} $ and $ c_{2} $ are integral constants and they are related by Eq.(23).\\
Now, we want to find these constants. By substituting Eq.(23) into Eq.(20) we have,
\begin{align}
&\theta = \frac{i}{2}\, \sqrt{3 + (3 c - 1)^2}\, (\ln r - c_{2}) = is, \\
&\cos\theta = \cos(is) = \cosh(s) = \frac{e^{s} + e^{-s}}{2}.
\end{align}
$e^{2s}$ is found by Eq.(27) as the following,
 \begin{equation}
e^{2s} = e^{-\sqrt{3 + (3 c - 1)^2}\, c_{2}}\,   r^{\sqrt{3 + (3 c - 1)^2}}.
\end{equation}
So the line element is as follows,
\begin{align}
ds^{2} =& r^{2 c} \Bigg(\frac{3 + (3 c - 1)^2}{12 \Lambda r^2} (\sqrt[3]{\frac{4}{e^{2s} + e^{-2s} + 2}})\Bigg) dt^{2} \nonumber\\& -\Bigg(\frac{3 + (3 c - 1)^2}{12 \Lambda r^2} \frac{4}{e^{2s} + e^{-2s} + 2} \Bigg)  dr^{2} \nonumber\\&
 - r^{- c} \Bigg(\frac{3 + (3 c - 1)^2}{12 \Lambda r^2}  \frac{4}{e^{2s} + e^{-2s} + 2}\Bigg)^{\frac{1}{3}} (r^2 d\varphi^{2} + dz^{2}).
\end{align}
If $c_2$ is real, the solution does not have an event horizon. Since, we want to find the black hole solution, we consider $c_2$ as a complex number and this condition guaranties $g^{rr}=0$.   
 \begin{equation}
c_{2} = c_{2}' + i c_{2}''.
\end{equation}
By inserting Eq.(31) in Eq.(27), we will have, 
 \begin{equation}
\theta = ( c_{2}' + a_{1} \ln r ) i + a_{2},
\end{equation}
where,
\begin{align}
& a_{1} = \frac{1}{2}\, \sqrt{3 + (3 c - 1)^2}\,  \ln r, \\
& a_{2} = - \frac{ c_{2}''}{2}\, \sqrt{3 + (3 c - 1)^2}.
\end{align}
Therefore, we have the following relations,
\begin{align}
&\cos\theta = \cos \Bigg(a_{2} + (c_{2}' + a_{1} \ln r) i \Bigg)\nonumber\\& = \cos a_{2}\, \cosh(c_{2}' + a_{1} \ln r) - i \sin a_{2}\, \sinh(c_{2}' + a_{1} \ln r),
\end{align}
\begin{align}
 &{\cos}^2\theta = {\cos}^2 a_{2}\, {\cosh}^2(c_{2}' + a_{1} \ln r) - {\sin}^2 a_{2}\, {\sinh}^2(c_{2}' + a_{1} \ln r)\nonumber\\&
-2 i\, \cos  a_{2}\, \sin a_{2}\, \cosh(c_{2}' + a_{1} \ln r)\, \sinh(c_{1}' + a_{1} \ln r).
\end{align}
Since the components of metric should be positive, the imaginary part of ${\cos}^2\theta$ should be zero. By applying this condition we will have,
 \begin{align}
\cos  a_{2} = 0 \qquad\textrm{or} \qquad \sin a_{2} = 0.
\end{align}
Now, we have two choices:\\
The first is as follows,
 \begin{equation}
\cos^2\theta = \cosh^2(c_{2}' + a_{1} \ln r) ,
\end{equation}
In this case, we don't have event horizons since $\cosh^2(c_{2}' + a_{1} \ln r)\neq 0$.\\
The second is as the following,
\begin{align}
&\cos^2\theta = -\sinh^2(c_{2}' + a_{1} \ln r),
\end{align}
In the above case, we have an event horizon because $\sinh^2(c_{2}' + a_{1} \ln r) = 0$. So we have,
 \begin{equation}\label{EH}
 r_+ = e^{-\frac{c_{2}'}{a_{1}}}.
\end{equation}
Finally, the metric of our model is as following,
\begin{align}\label{solution}
 ds^{2} =& r^{2 c} \Bigg(\frac{3 + (3 c - 1)^2}{12 \Lambda r^2 (-{\sinh}^2(c_{2}' + a_{1} \ln r) )} \Bigg) dt^{2} -  \Bigg(\frac{3 + (3 c - 1)^2}{12 \Lambda r^2 (-{\sinh}^2(c_{2}' + a_{1} \ln r))}  \Bigg)  dr^{2} \nonumber\\&
  - r^{- c} {\Bigg(\frac{3 + (3 c - 1)^2}{12 \Lambda r^2 (-{\sinh}^2(c_{2}' + a_{1} \ln r))} \Bigg)}^{\frac{1}{3}} (r^2 d\varphi^{2} + dz^{2}).
\end{align}
Black hole can be considered as a thermodynamic system since temperature and entropy are assigned to it. The effect of quantum field on the classical geometry can be revealed by calculation of thermodynamic quantities \cite{Hayward:1999ek}-\cite{Mistry:2017ubm}. The temperature \cite{Hawking:1976de} is as following,
 \begin{align}
	T =\frac{1}{2 \pi}\Bigg[ \frac{1}{\sqrt{-g_{rr}}} \frac{d}{dr}\sqrt{g_{tt}}\Bigg]_{r=r_+}=\frac{ 4 c c'_2 +2 a_1 (-1+ 2c)\ln r_+-a_1- 2c'_2}{4 \pi r_+^{1-c}  (c_{2}'+a_1 \ln r_+)}.
\end{align}
To have a finite temperature, we choose $a_1=0$. So, by using Eq.(\ref{EH}) event horizon is located on the infinity.
 \begin{equation}
	T =\frac{r_+^{c-1}}{4 \pi}(2+4c ).
\end{equation}
The temperature is very large for  $c \geq 1$ and the temperature tends to zero for  $c < 1$. For $c > \frac{-1}{2}$, we have positive temperature.\\
The temperature of the black hole is plotted for $c=\frac{4}{3}$ in Fig \ref{fig:1}.\\
\\
The Bekenstein–Hawking entropy \cite{Bekenstein:1973ur} is as follows,
 \begin{align}\label{entropy1}
	S& =\frac{A}{4 G}= \frac{1}{4 G}\int d^2x \sqrt{g} = \frac{V_2}{4 r_+^{ c-1} G}  {\Bigg(\frac{3 + (3 c - 1)^2}{-12 \Lambda r_+^2 {\sinh}^2(c_{2}' + a_{1} \ln r_+)} \Bigg)}^{\frac{1}{3}}. 
\end{align} 
where,
 \begin{equation}
	V_2 =2 \pi l r_+^{1-c}{\Bigg(\frac{3 + (3 c - 1)^2}{-12 \Lambda r_+^2 {\sinh}^2(c_{2}' + a_{1} \ln r_+)} \Bigg)}^{\frac{1}{3}},
\end{equation}
by substituting $V_2$ in Eq.(\ref{entropy1}), we have,\\
 \begin{align}\label{entropy}
	S& = \frac{2 \pi l}{4 r_+^{2 c-2} G}  {\Bigg(\frac{3 + (3 c - 1)^2}{-12 \Lambda r_+^2 {\sinh}^2(c_{2}' + a_{1} \ln r_+)} \Bigg)}^{\frac{2}{3}}. 
\end{align} 
The mass of the black hole is found by Smarr relation\cite{Smarr:1972kt}, 
 \begin{equation}
	M =2(TS-PV),
\end{equation}
the cosmological constant plays as a pressure, $P=-\frac{\Lambda}{8 \pi G}$. So the mass is as the following,
\begin{align}
	M&=2\pi l (1+2c ) r_+^{-c}  {\Bigg(\frac{3 + (3 c - 1)^2}{-12 \Lambda r_+^2 {\sinh}^2(c_{2}' + a_{1} \ln r_+)} \Bigg)}^{\frac{2}{3}}\nonumber\\&-\frac{3r_+^{1-c}}{2 G l}{\Bigg(\frac{3 + (3 c - 1)^2}{-12 \Lambda r_+^2 {\sinh}^2(c_{2}' + a_{1} \ln r_+)} \Bigg)}^{\frac{1}{3}}.
\end{align}
To have finite mass, the following relation should be satisfied,
\begin{equation}\label{cond1}
	3+(3c-1)^2={\sinh}^2(c_{2}' + a_{1} \ln r).
\end{equation}
Therefore, the mass is given by,
\begin{equation}
M=\frac{(4 c+2) l^{\frac{7}{3}} }{8 \sqrt[3]{6} G  r^{c+\frac{1}{3}}}-\frac{\sqrt[3]{3} }{2\ 2^{2/3} G \sqrt[3]{l}r^{c-\frac{1}{3}}}.
\end{equation}
The energy of the black hole is found by using the first law of  thermodynamics \cite{Kichakova:2015nni},\cite{Delgado:2022pcc} as following,
\begin{equation}
dE=TdS+PdV.
\end{equation}
It can be written as,
\begin{equation}
	dE=T(\frac{\partial S}{\partial r_+})d r_++P(\frac{\partial V}{\partial r_+})d r_+.
\end{equation}
Energy of this black hole is found by integrating with respct to $r_+$,
\begin{equation}
	E=\frac{(4 c+2) l^{\frac{7}{3}} }{8 G r_+^{c+\frac{1}{3}} }+\frac{\sqrt[3]{3} }{4\ 2^{\frac{7}{3}} G l^{\frac{1}{3}} r_+^{c-\frac{1}{3}}},
\end{equation}
The first law of thermodynamics is satisfied. Our solution Eq.(\ref{solution}) is asymptotically AdS  by applying Eq.(\ref{cond1}) and considering $c=\frac{4}{3}$.  \\
We will study the Hawking-Page phase transition or the first order phase transition \cite{Ranjbari:2019ktp}-\cite{Ghanaatian:2019xhi} in our future work.

 \section{Results and discussion}
We solved the Einstein's equations for a mass line in the 4-dimensional space-time with negative cosmological constant. The solution of Einstein's equations with event horizon is called black hole. In this paper, we introduced the black hole solution with cylindrical symmetry in AdS space-time. The field theory dual of this model will be studied by using the fluid-gravity duality \cite{Policastro2001}-\cite{Sadeghi:2021qou} in our future work. 

 \section{Conclusion}
\noindent In this paper, we studied the static cylindrically symmetric vacuum solutions of the  Einstein equations in AdS space-time. Also, we introduced the cylindrical black hole or the black string of this model. We obtained the first law of thermodynamics for this black string by introducing temperature, Bekenstein-Hawking entropy and pressure. Finally, we mention some future directions of this model.\\
 One of the key challenges in the study of static cylindrically symmetric solutions is determining their stability under perturbations. Future works can focus on analyzing the stability of these solutions and investigating the conditions under which they are stable.\\
The study of holographic aspects of this solution is an interesting subject. The AdS/CFT correspondence is a powerful tool to study the relationship between gravitational theories and quantum field theories. Using holographic techniques to study static cylindrical solutions in AdS space-time and exploring their connection to other areas of physics is an excellent subject.\\
Since most of the existing literature has focused on cylindrical solutions in AdS space-time, the extension of cylindrical solutions to other space-times is an untouched subject.\\
Another promising direction for future research is the study of cylindrical solutions in higher-dimensional space-times. This could provide insights into the behavior of matter and energy in higher-dimensional space-times and potentially lead to new discoveries in the field of high energy physics.\\
Therefore, the study of static cylindrically symmetric vacuum solutions of the Einstein equations in AdS space-time is an active area of research with many open questions and potential ways for future exploration.

\vspace{1cm}
\noindent {\large {\bf Acknowledgment} } Authors gratefully thank to Ahmad Moradpouri, Komeil Babaei, Faramarz Rahmani, Ruben Campos Delgado and the referee of Indian Journal of Physics for the useful comments and suggestions which definitely help to improve the manuscript.



\begin{thebibliography}{}
	
\bibitem{Einstein:1914bt}
A~Einstein
\textit{Phys. Z. } \textbf{15} (1914)
	
\bibitem{Einstein:1914bx}
A ~Einstein
\textit{Sitzungsber. Preuss. Akad. Wiss. Berlin (Math. Phys. )} (1914)
	
\bibitem{Einstein:1915by}
A ~Einstein
\textit{Sitzungsber. Preuss. Akad. Wiss. Berlin (Math. Phys. )} (1915)	


\bibitem{deRham:2010kj}
C~de Rham, G~Gabadadze and A~J~Tolley
\textit{Phys. Rev. Lett.} \textbf{106} 231101 (2011)

	
\bibitem{Sadeghi:2015vaa}
M ~Sadeghi and S ~Parvizi
\textit{Class. Quant. Grav.} \textbf{33} 035005 (2016)


\bibitem{Parvizi:2017boc}
S~Parvizi and M~Sadeghi
\textit{Eur. Phys. J. C}  \textbf{79} 113 (2019)


	

	



\bibitem{Sadeghi:2018vrf}
M ~Sadeghi
\textit{Mod. Phys. Lett. A }\textbf{33} 1850220 (2018)


\bibitem{Levi-Civita}
Levi-Civita, T. 1919 \textit{Rend. Acc. Lincei } \textbf{36} 3625 (1995)

\bibitem{Weyl:1917gp}
H ~Weyl
Annalen Phys. \textbf{54} 117 (1917)


\bibitem{Lemos:1994xp}
J~P~S~Lemos
\textit{Phys. Lett. B} \textbf{353} 46 (1995)

\bibitem{Lemos:1997bd}
J~P~S~Lemos
\textit{Phys. Rev. D} \textbf{57} 4600 (1998)


	\bibitem{Alexander Vilenkin}
	A~ Vilenkin
	\textit{Physics Reports } \textbf{121} 263 (1985)



\bibitem{Bronnikov:2019clf}
K~Bronnikov, N~O~Santos and A~Wang
\textit{Class. Quant. Grav.} \textbf{37} 113002 (2020)



\bibitem{Hayward:1999ek}
S~A~Hayward
\textit{Class. Quant. Grav.} \textbf{17} 1749 (2000)

\bibitem{Kichakova:2015nni}
O~Kichakova, J~Kunz, E~Radu and Y~Shnir
\textit{Phys. Rev. D} \textbf{93}  044037 (2016)

\bibitem{Mistry:2017ubm}
R~Mistry, S~Upadhyay, A~F~Ali and M~Faizal
\textit{Nucl. Phys. B} \textbf{923} 378 (2017)


\bibitem{Hawking:1976de}
S~W~Hawking
\textit{Phys. Rev. D} \textbf{13} 191 (1976)


\bibitem{Bekenstein:1973ur}
J D Bekenstein 
\textit{Phys. Rev. D} \textbf{7} 2333 (1973)



\bibitem{Smarr:1972kt}
Smarr L~
\textit{Phys. Rev. Lett.} \textbf{30} 71 (1973)


\bibitem{Delgado:2022pcc}
R~C~Delgado
\textit{Eur. Phys. J. C} \textbf{82} 272 (2022)


\bibitem{Ranjbari:2019ktp}
H~Ranjbari, M~Sadeghi, M~Ghanaatian and G~Forozani
\textit{Eur. Phys. J. C} \textbf{80} 17 (2020)




\bibitem{Ghanaatian:2019xhi}
M~Ghanaatian, M~Sadeghi, H~Ranjbari and G~Forozani
\textit{Mod. Phys. Lett. A} \textbf{35} 2050203 (2020)






\bibitem{Policastro2001}
G~Policastro, D~T~Son and A~O~Starinets
\textit{Phys.\ Rev.\ Lett.\ } {\bf 87} 081601 (2001)




\bibitem{Sadeghi:2018ylh}
M~Sadeghi
\textit{Eur. Phys. J. C} \textbf{78} 875 (2018)







\bibitem{Sadeghi:2019trg}
M~Sadeghi
\textit{Indian J. Phys.} \textbf{94} 1119 (2020)

\bibitem{Sadeghi:2021qou}
M~Sadeghi
\textit{Indian J. Phys.} \textbf{96} 4341 (2022)


\end{thebibliography}
\end{document}